\documentclass[namedreferences,hyperref,optionalrh,solaromanenum]{spr-sola}

\usepackage{graphicx}                    % For eps figures, newer & more powerfull
\usepackage{amssymb}       % useful mathematical symbols
\usepackage{color}     
\usepackage{makecell}
\usepackage{booktabs}
\usepackage{amsmath}
\usepackage{multirow}
\newcommand{\aap}{{\it Astron. Astrophys.}}

\newcommand{\apjl}{{\it Astrophys. J. Lett.}}

\newcommand{\solphys}{{\it Sol. Phys.}}
 
\newcommand{\ssr}{{\it Space Sci. Rev.}} 
\chardef\us=`\_

\defcitealias{Lu1993}{LH93}
\defcitealias{Lu1991}{LH91}

\begin{document}

%Flare Cascades: Avalanche Models of Sympathetic Solar Flares
\begin{frontmatter}

\title{Avalanching together: A model for sympathetic flaring}

%%%%%%%%%%%%%%%%%%%%%%%%%%%%%%%%%%%%%%%%%%%%%%%%%%%
%% Authors Names
%
\author[addressref={aff1},corref, email={louis-simon.guite@umontreal.ca}]{\inits{L-S. }\fnm{Louis-Simon~}\snm{Guité}}
\author[addressref={aff1},email={paul.charbonneau@umontreal.ca}]{\inits{P. }\fnm{Paul~}\snm{Charbonneau}}
\author[addressref={aff2},email={antoine.strugarek@cea.fr}]{\inits{A. }\fnm{Antoine~}\snm{Strugarek}}
%%%%%%%%%%%%%%%%%%%%%%%%%%%%%%%%%%%%%%%%%%%%%%%%%%%
%% Runningheads
%
\runningauthor{Guité, Charbonneau and Strugarek}
\runningtitle{Avalanching together: A model for sympathetic flaring}

%%%%%%%%%%%%%%%%%%%%%%%%%%%%%%%%%%%%%%%%%%%%%%%%%%%
%% Affilations 
%% id shold be the same with \author addressref value.
\address[id={aff1}]{Physics Department, Université de Montréal, CP 6128 Centre-Ville, Montréal, QC H3C-3J7, Canada}
\address[id={aff2}]{Université Paris-Saclay, Université Paris Cité, CEA, CNRS, AIM, 91191, Gif-sur-Yvette, France}
%%%%%%%%%%%%%%%%%%%%%%%%%%%%%%%%%%%%%%%%%%%%%%%%%%%
%%% Abstract 
\begin{abstract}
Avalanche models running in a self-organized critical regime have proven powerful in reproducing the power-law distributions and scale invariance that characterize the statistical properties of solar flares. They are often interpreted as representing an individual active region of the Sun. As a result, this class of models has rarely been applied to describe sympathetic flares\textemdash solar eruptions that occur in close spatial and temporal proximity, seemingly driven by their mutual interaction. In this study, we investigate the phenomenon of sympathetic flaring using avalanche models and compare their statistical properties with observations of sympathetic flares on the Sun. We developed a novel avalanche model featuring two connected lattices, each representing a distinct active region. This connectivity allows the transfer of nodal variable between the lattices, simulating the non-local effects expected to occur during sympathetic flares. Our results show that under strong connectivity, the lattices exhibit temporal synchronization, with correlations between their avalanche energies. Furthermore, increasing the connectivity between the lattices results in an excess of avalanches at short waiting times. A quantitative comparison with observational data suggests that only a weak connectivity allows our model to replicate the observed solar waiting time distributions. Consequently, we propose that if magnetic connectivity between distinct active regions drives sympathetic flaring on the Sun, it must remain relatively weak.

\end{abstract}

%%%%%%%%%%%%%%%%%%%%%%%%%%%%%%%%%%%%%%%%%%%%%%%%%%%
%% Keywords
%
\keywords{Avalanche Models - Sympathetic Solar Flares}

\end{frontmatter}
%-------------------------------------------------

%%%%%%%%%%%%%%%%%%%%%%%%%%%%%%%%%%%%%%%%%%%%%%%%%%%
%% Sections
%

\section{Introduction}\label{sec:introduction}

Solar flare observations at short wavelengths have revealed very well-defined power-law frequency
distributions in many flare size measures such as peak flux, duration, and total released energy
\citep[see, e.g.,][and references therein]{Dennis1985, Aschwanden2016}.
This is usually taken to indicate
scale invariance in the flaring process, a property shared by many other natural systems accumulating energy slowly, but releasing it rapidly in a spatiotemporally intermittent manner; in this respect, and despite widely varying scales and underlying physical mechanisms,
flares are akin to earthquakes, landslides, avalanches, floods and forest fires.

Self-organized criticality \citep[hereafter SOC; see, e.g.,][]{Bak1987, Jensen1998, Aschwanden2016} has emerged
as a robust generator of such scale-invariant energy release behavior in natural systems.
Under this view, flares of all sizes
are the global result of an ``avalanche'' of small-scale elementary reconnection events
occurring within a stressed coronal magnetic structure, such as magnetic loops overlying active regions.
Under this Ansatz, large flares only differ from small ones in the greater number of magnetically stressed sites undergoing reconnection \citep{Lu1993,Lu1995}.
Such elementary reconnection events could occur
at sites of magnetic tangential discontinuities developing naturally in low-$\beta$ stressed magnetic structures, as described for example by \cite{Parker1988}.

The class of cellular automata known as sandpile (or avalanche) models has been used extensively
to model SOC dynamics in natural systems, including in the flare context \citep[e.g.,][]{Lu1993, Georgoulis2001, Charbonneau2001,  Strugarek2014, Thibeault2022, Lamarre2024}. Coronal magnetic structures are mapped to a lattice of interconnected nodes undergoing slow energy loading and subject to a self-limiting local threshold instability.
Loading, instability threshold, and local redistribution of nodal variable from unstable nodes are all modeled using very simple nodal update rules, inspired from
reconnection physics to the extent possible (more on this in \S 2 below).
As abstract as they may be, at this writing such models remain the most robust generators
for the power-law form observed in the frequency
distribution of solar (and stellar) flare size measures.

Under statistically stationary loading, SOC models typically exhibit an exponential distribution 
of inter-event waiting times (WTD), as one would expect from a memoryless random process 
\citep{Wheatland2001,Norman2001}.
Sympathetic flaring, namely the occurrence of pairs of flares closely spaced in time
but occurring in distinct active regions, leads to an excess of short waiting times \citep[e.g.,][and references therein]{Fritzova-Svestkova1976, Pearce1990, Moon2002, Schrijver2015, Guité2025}. The sympathetic flaring phenomenon poses a challenge to the classical SOC picture,
which we take up in this paper.

We begin in Section 2 with a brief description of the SOC lattice model of
Lu \& Hamilton \citep[see][hereafter \citetalias{Lu1991} and \citetalias{Lu1993} respectively]{Lu1991, Lu1993}, along with a more detailed description of the lattice connectivity scheme designed for the
present study.
In Section 3 we investigate the conditions under which connectivity can lead to sympathetic avalanching, the model's equivalent to sympathetic flaring,
and how the waiting time distribution for the collective avalanching behavior
of the connected lattices is affected by the strength of the connectivity.
We also examine under which conditions the SOC state can be sustained under connectivity,
and how the latter can produce (or not) statistically significant correlations between the energies of sympathetic flare/avalanche pairs.
In Section 4 we use the sympathetic flare database assembled by \cite{Guité2025} to identify
the regions of the model's parameter space that are consistent with observations,
thus allowing us to place constraints on the acceptable level of lattice connectivity representative of the Sun's sympathetic flaring behavior.
We conclude in Section 5 by summarizing our modeling results and
briefly discussing their implications for observations of sympathetic stellar flares, particularly for stars with much stronger magnetic fields than the Sun.

\section{Sandpile Models of Solar Flares}\label{sec:sandpile}

\subsection{The Lu \& Hamilton Model}\label{sec:LH}
We focus on the 2-dimensional variant of the \citetalias{Lu1991} model (see \citealt{Charbonneau2001} for a comprehensive review), where each node in a lattice of linear size $N$ is assigned a continuous scalar value, as described in \citetalias{Lu1993}. This simplification is justified by findings that 2D and 3D lattices exhibit similar statistical properties \citep{Charbonneau2001}, and scalar-field models yield power-law distributions comparable to those of vector-field models \citep{Robinson1994}. 
\par In the \citetalias{Lu1993} model formulation, the nodal variable is $A^{t}_{i,j}$, where $A$ represents the z-component of the magnetic vector potential, ($i,j$) denotes the node's position in the lattice, and $t$ refers to the temporal iteration of the simulation. From a given initial condition, usually chosen to be $A = 0$ everywhere, this cellular automaton is driven by adding a random increment $\delta A$ to a randomly chosen interior node (changing from one iteration to the next) while maintaining the boundary nodes at $A = 0$. This increment is drawn from a uniform distribution $\delta A \in [\sigma_{1}, \sigma_{2}]$, where $\sigma_{1}$ and $\sigma_{2}$ determine the amount of driving on the system. These parameters are chosen so that $\langle \delta A \rangle > 0$ to allow the growth of a mean field on the lattice. Furthermore, the condition of weak driving $\delta A / \langle A \rangle \ll 1$ is crucial to reach a state of SOC. One can define the curvature at a particular node to be
    \begin{equation}
        \Delta A_{i,j}^{t} \equiv A_{i,j}^{t} - \frac{1}{4}\sum_{n=1}^{4}A_{n}^{t}\,,
    \end{equation}
    
    where $n$ corresponds to the neighboring nodes $(i\pm 1,j)$ and $(i, j\pm 1)$. At each temporal iteration, the stability of every interior node is evaluated by comparing its curvature to a predetermined threshold value, denoted as $Z_{c}$. If any $(i,j)$ node satisfies $|\Delta A_{i,j}^{t}| > Z_{c}$, the node is deemed unstable and will trigger a local redistribution of nodal variable in order to restore stability, under the following redistribution rules,
\begin{align}\label{eqn:redistribution}
        A_{i,j}^{t+1} &=  A_{i,j}^{t} - \frac{4}{5}Z \hspace{0.4cm} (\textrm{Unstable node}) \\
        A_{n}^{t+1} &= A_{n}^{t} + \frac{1}{5}Z \hspace{0.54cm} (\textrm{Neighboring nodes})
    \end{align}
    
    where $Z \equiv Z_{c} \cdot\, \operatorname{sign(\Delta A)}$. Following this isotropic redistribution, one of the neighboring nodes might become unstable, leading to a cascade effect that propagates the avalanche throughout the lattice until stability is restored everywhere in the system. All the avalanching nodes are updated simultaneously in the lattice to avoid introducing a spatial bias related to the order in which the curvature stability is checked. It is worth noting that these redistribution rules conserve the sum of nodal variable ($\sum_{i,j}A_{i,j}$) in time, except when an avalanche reaches the lattice boundaries, where the nodal variable is set to zero. If we define the energy of a node to be $A_{i,j}^{2}$, then the redistribution reduces the energy of the unstable node and its neighbors by an amount \begin{equation}\label{eqn:Eredistribution}
        \Delta e_{i,j}^{t} = \frac{4}{5}\,Z_{c}^{2}\,\left(\frac{2\Delta A_{i,j}^{t}}{Z} -1 \right).
    \end{equation}
    
    Although $A^{2}$ does not directly measure magnetic energy like $B^{2}$ in the original \citetalias{Lu1991} model, the interpretation of $A_{i,j} \equiv A_{z}$ (z-component of the magnetic vector potential) as being the amount of twist of a coronal loop's magnetic field justifies its adoption as a magnetic energy definition, as shown in \cite{Lamarre2024}. If the curvature stability threshold is exceeded by an infinitesimal amount (i.e., $|\Delta A_{i,j}| \approx Z_{c}$), the energy release is expressed as
    \begin{equation}\label{eqn:quantum}
        e_{0} = \frac{4Z_{c}^{2}}{5}.
    \end{equation}
    
    This energy quantum serves as a convenient unit for describing the energy of the system.
    
    \subsection{Connected Lattices Model}\label{sec:coupled}
        
    The phenomenon of sympathetic solar flares has been explored within the framework of avalanche \citep{MacKinnon1997} and separator \citep{Wheatland2006} models. In these studies, the authors introduced an increase in the probability of avalanches occurring beyond neighboring nodes following the onset of a flare. However, these non-local effects are only considered within a single lattice in this case. If we assume that a lattice can simplistically represent an active region, it would be more relevant to model the non-local effects between two distinct lattices, in order to capture the phenomenon of sympathetic flares between different active regions. We therefore propose a novel avalanche model consisting of two interacting lattices that replicate the dynamics observed in sympathetic flaring. This lattice connectivity is motivated by the idea that magnetic field lines linking different active regions may allow the propagation of information between them, thus enabling the trigger of sympathetic flares \citep{Guité2025}.

    \subsubsection{Connectivity methodology}
    
    \par
\begin{figure}[h!]
            \centering
            \includegraphics[width=\textwidth]{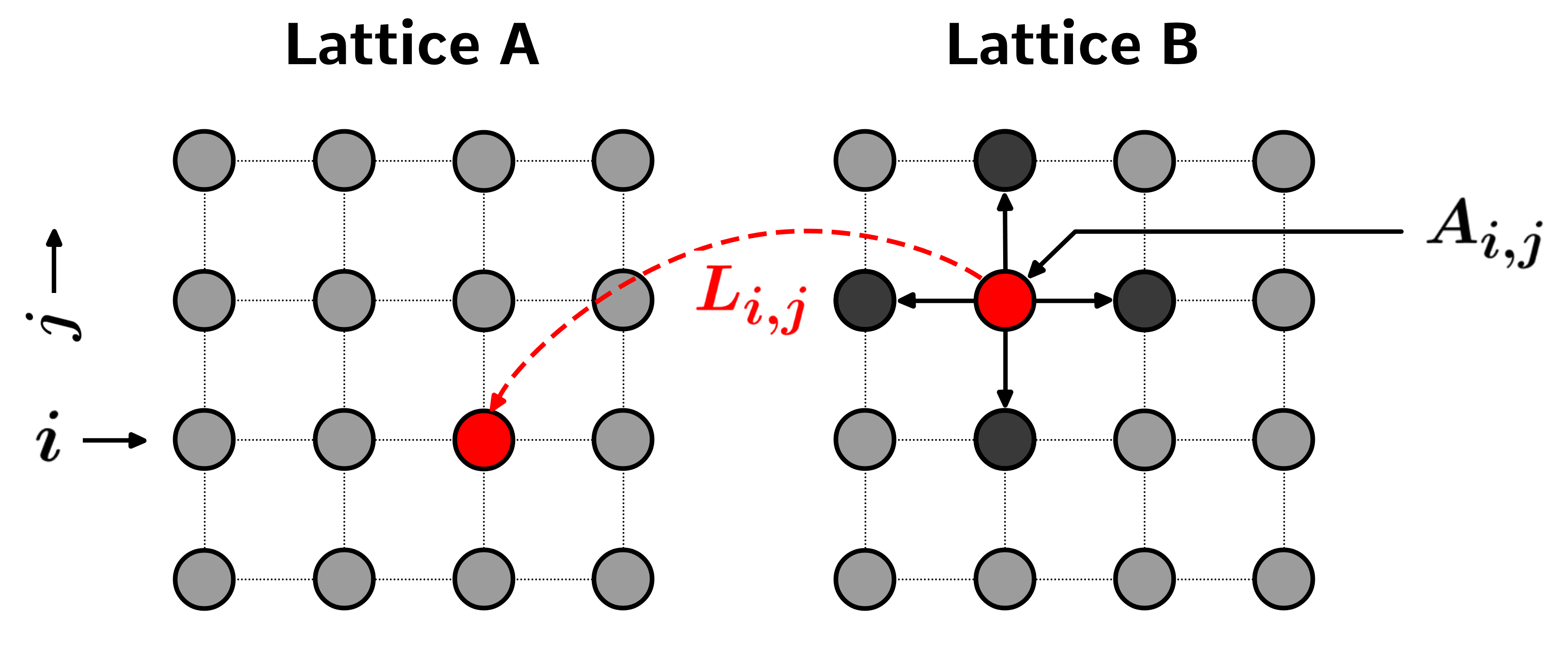}
            \caption{Schematic of connected lattices, where red nodes are connected together. Arrows show the direction of redistribution for an avalanching node (details in the text). $A_{i,j}$ is the nodal variable and $L_{i,j}$ is the amount transferred given by Equation (\ref{eqn:nodal_transfer}). Note that the connection is bidirectional, meaning that nodal variable can be transferred both ways.}
            \label{fig:schematic}
        \end{figure}

 The procedure to establish a connection between the lattices is as follows. First, we introduce a parameter $f$ from the interval $f \in [0,1]$, representing the fraction of total nodes in each lattice that are connected. Second, we choose $fN^{2}$ distinct, random pairs of connected nodes between the two lattices, ensuring that boundary nodes are excluded and that each node can be connected to only one node in the other lattice. This connectivity enables the transfer of nodal variable between lattices when a connected node undergoes an avalanche. The non-conservative redistribution rules for connected nodes are given by
    \begin{align}
            A_{i,j}^{t+1} &=  A_{i,j}^{t} - \frac{4Z}{5}\hspace{1.03cm}(\textrm{Unstable node}) \label{eqn:NC_redistribution1} \\
            A_{n}^{t+1} &= A_{n}^{t} + \frac{(1-\xi)Z}{5} \hspace{0.3cm}(\textrm{Neighboring nodes}) \label{eqn:NC_redistribution2}
    \end{align}
    where $\xi$ is uniformly drawn from a distribution $\xi \in [0,\, \alpha]$, with $\alpha \in [0,\,1]$ being a non-conservative transfer parameter to be specified for the simulation. In this way, $\alpha = 0$ corresponds to a completely conservative model with no transfer between the lattices. The amount of nodal variable that is transferred to the connected node in the other lattice is then given by 
    \begin{equation}
    \label{eqn:nodal_transfer}
        L^{t}_{i,j} = \frac{4}{5}Z_{c}\,\xi,
    \end{equation}
    which represents the portion not redistributed to neighboring nodes. Due to the symmetry of the connections, a lattice will, on average, lose as much nodal variable as it gains. For unconnected nodes, conservative redistribution rules ensure that the total sum of both lattices remains conserved. We also adopt a random driving mechanism similar to that in \citetalias{Lu1993}. A schematic of this connectivity is illustrated in Figure \ref{fig:schematic}. The red node in lattice B is unstable and redistributes nodal variable to its four nearest neighbors (dark gray nodes) according to Equations (\ref{eqn:NC_redistribution1}) and (\ref{eqn:NC_redistribution2}). Additionally, it transfers an amount specified by Equation (\ref{eqn:nodal_transfer}) to its connected node in lattice A (also highlighted in red).

        \subsubsection{Running sandpile model}
        \par Avalanche models of solar flares typically pause external driving during an avalanche (referred as \textit{Stop-and-Go} hereafter) to account for the timescale separation between driving and avalanching processes ($t_{\textrm{avalanche}} \ll t_{\textrm{driving}}$). However, with the inclusion of two lattices, it is important to carefully manage the driving to ensure that both lattices experience the same elapsed simulation time. This can be achieved in two ways. The first approach involves pausing the driving of a lattice that is not avalanching if the other lattice is. The paused iterations are flagged to prevent them from contributing to the total elapsed time of the inactive lattice. For comparison between the lattices, the time series must then be compressed by combining the avalanching iterations into a single iteration. This process results in the loss of intra-avalanche dynamics, during which transfer between lattices can occur\textemdash an aspect that is central to the study. Thus, we do not consider this method. 
        \par
\begin{figure}[h!]
            \centering
            \includegraphics[width = \textwidth]{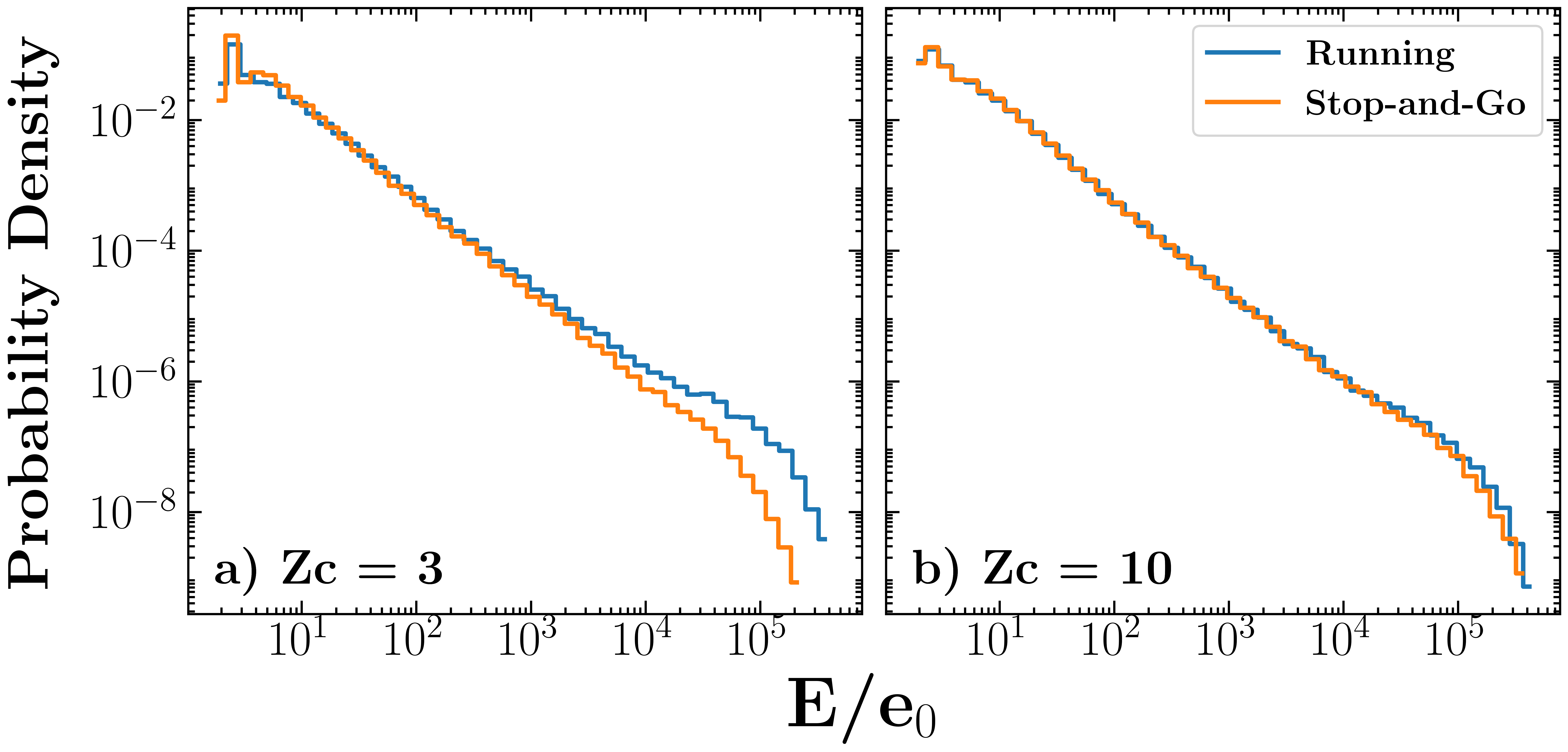}
            \caption{Distribution of avalanche total energy for $Z_{c} = 3$ (a) and $Z_{c} = 10$ (b), with orange being Stop-and-Go and blue running sandpile model. We consider a single lattice of linear size $N$ = 48 with a \citetalias{Lu1993}  model, and $\delta A \in [-0.2, 0.8]$.}
        \label{fig:running_sandpile}
        \end{figure}

 The alternative approach is to continuously drive both lattices at each iteration, regardless of whether they are avalanching. This is known as a \textit{running sandpile model} \citep{HwaKardar89,Hwaetal92}. If the driving is weak ($\delta A \ll Z_{c}$), the avalanche dynamics should remain largely unaffected by the constant driving. This is presented in Figure \ref{fig:running_sandpile}, where we show the distributions of total avalanche energy for $Z_{c} = 3$ (a) and $Z_{c} = 10$ (b) for a single lattice of linear size $N$ = 48 with a \citetalias{Lu1993} model, and $\delta A \in [-0.2, 0.8]$. The orange distribution is for the typical Stop-and-Go driving, while the blue is for the running sandpile model. For $Z_{c} = 3$, the distributions differ for high energies, whereas they are identical for $Z_{c} = 10$. This indicates that the energy dissipation timescale during an avalanche remains significantly shorter than the driving timescale, even under constant driving, ensuring the regime $t_{\textrm{avalanche}} \ll t_{\textrm{driving}}$. This validates the use of a running sandpile model for the connected lattices, provided a high critical threshold value, such as $Z_{c} \gg \max(|\delta A|)$, is selected.
        The choice $Z_c=10$ will be maintained throughout the analysis, significantly simplifying the process by ensuring the time series remain temporally aligned. Similar results were observed for the peak energy ($P$) and avalanche duration ($T$) distributions, although these are not shown here for conciseness. 

        Key characteristics of avalanche models are evident in the distributions shown in Figure \ref{fig:running_sandpile}. First, the distributions follow a power-law function, highlighting the scale invariance of the model when the lattice has reached the SOC state. Second, the distributions taper off at high energies, a consequence of the lattice's finite size, which limits the extent of avalanche propagation and the maximum energy release. Lastly, the distributions flatten at low energies due to the lattice's granularity at small scales, as there exists a minimum energy that can be released by a single avalanche.
                
\section{Properties of connected sandpile models}\label{sec:results}

        In this section, we consider 9 main simulations combining $f$ values of 0.01, 0.1, and 0.25 
        with $\alpha$ values of 0.02, 0.2, and 0.9, all with linear size $N$ = 48, $Z_{c} = 10$, and random driving $\delta A \in [-0.2,\, 0.8]$ (as in \citetalias{Lu1991}).

    \subsection{Waiting Time Distribution}\label{sec:waiting_time}

\begin{figure}[h!]
        \centering
        \includegraphics[width=\textwidth]{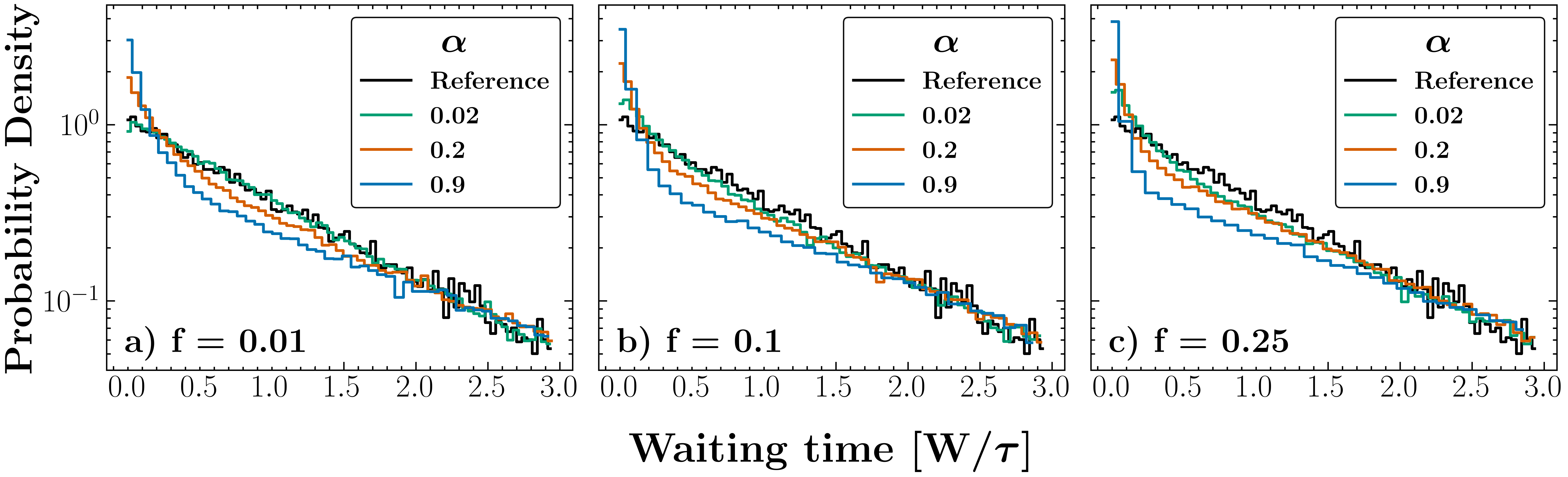}
        \caption{Waiting time distributions for $f = 0.01$ (a), $f = 0.1$ (b), and $f = 0.25$ (c), with $\alpha$ values of 0.02 (green), 0.2 (orange), and 0.9 (blue). The black distribution represents a reference simulation, where no transfer occurs between the lattices. Lattices have linear size $N = 48$ and $Z_c = 10$. The distributions are normalized by their arithmetic mean waiting time $\tau$.}
        \label{fig:W}
    \end{figure}

    The waiting time, denoted $W$, is defined as the number of iterations between two subsequent avalanches. In Figure \ref{fig:W}, we show the waiting time distributions for $f = 0.01$ (a), $f = 0.1$ (b), and $f = 0.25$ (c), with $\alpha$ values of 0.02 (green), 0.2 (orange), and 0.9 (blue). Since the two lattices are statistically identical, we combine their waiting time distributions to increase the event count. The black distribution represents a reference simulation, where no transfer occurs between the lattices. The waiting times are normalized by their corresponding arithmetic mean waiting time $\tau$ to allow a better comparison between simulations. 

\begin{figure}[H]
        \centering
    \includegraphics[width=\textwidth]{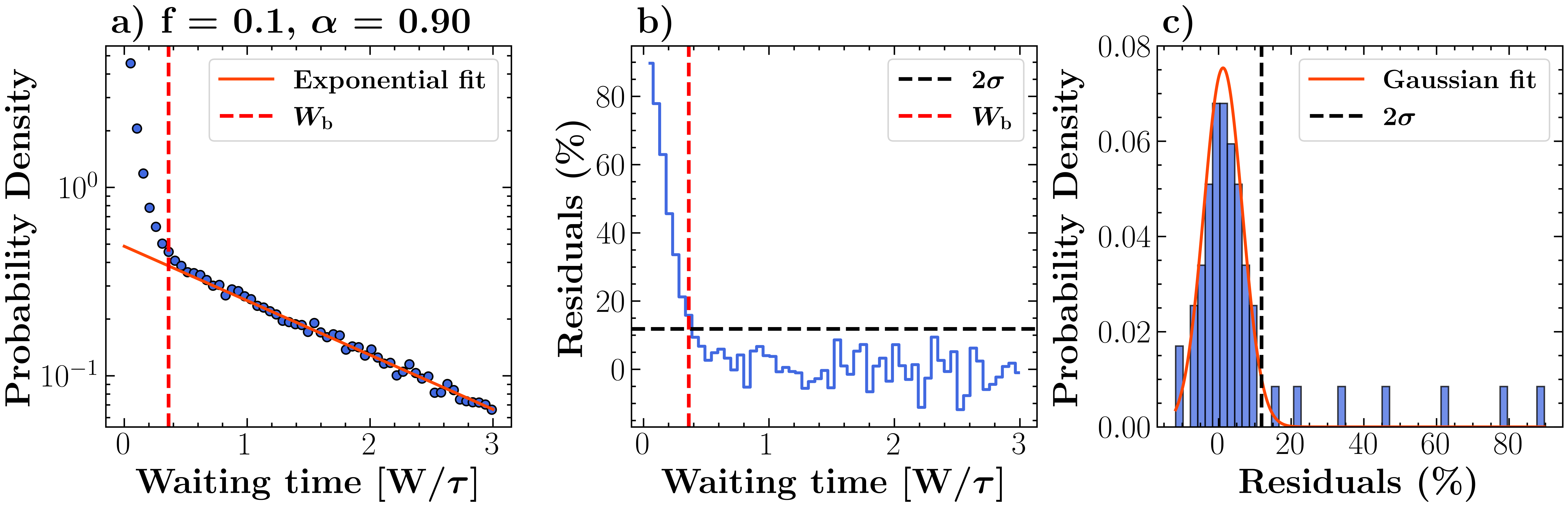}
        \caption{Waiting time distribution with an exponential fit (a), residuals as a function of waiting time (b), and the fit residuals histogram (c) for a simulation with $f = 0.1$ and $\alpha = 0.9$. The vertical dashed red line indicates the waiting time break $W_{\textrm{b}}$ while the dashed black line marks the $2\sigma$ threshold from the Gaussian fit.}
        \label{fig:exponential_fit}
    \end{figure}

    For a fixed number of connected nodes, increasing $\alpha$—which increases the transfer of nodal variable between lattices—causes the waiting time distribution to deviate from a simple exponential at shorter waiting times. We outline below the method used to better quantify the waiting time at which this break occurs, denoted as $W_{\textrm{b}}$. First, we fit a decaying exponential function to the tail of the waiting time distribution ($W/\tau > 1$), as shown by the solid red line in the panel (a) of Figure \ref{fig:exponential_fit}. Next, we compute the residuals between the exponential fit and the distribution as a function of the waiting time, shown in panel (b). To determine $W_{\textrm{b}}$, we plot the histogram of residuals and identify the waiting time corresponding to the $2\sigma$ threshold (vertical dashed black line) from a Gaussian fit (solid red line), as shown in panel (c). The value of $W_{\textrm{b}}$ is highlighted by the dashed red line in panels (a) and (b). In Figure \ref{fig:exponential_fit}, using parameters $f = 0.1$ and $\alpha = 0.9$, the break is determined to be $W_{\textrm{b}} \approx 0.36 \,\tau$, which is reasonable based on a visual inspection of panel (a). The values of $W_{\textrm{b}}$ for the nine simulations of Section \ref{sec:results} are listed in Table \ref{tab:metrics_table}. In Section \ref{sec:comparison_sdo}, we present the value of $W_{\textrm{b}}$ for a larger range of simulation parameters, in order to compare with real observations of solar flares.

    As shown by the black curve in Figure \ref{fig:W}, disconnecting the lattices and allowing them evolve independently causes the waiting time distribution to revert to an exponential form. Additionally, as the number of connected nodes increases (higher $f$, rightmost panel), this deviation becomes more significant. Furthermore, as shown in the second column of Table \ref{tab:metrics_table}, the mean waiting time $\tau$ decreases with increasing connectivity, indicating a higher frequency of avalanches. In summary, stronger lattice connectivity results in more frequent avalanches occurring over shorter waiting times. This stands in contrast to the typical \citetalias{Lu1993} model, which exhibits an exponential distribution across all values of $W$ for statistically stationary driving \citep[e.g.,][]{Charbonneau2001}.

 \vspace{0.5cm}
    \begin{table}[H]
    \begin{tabular}{c|c|c|c|c}
        Simulation parameters & $\tau$ & $W_{\textrm{b}}/\tau$ & r & Time lag \\       
        \Xhline{3\arrayrulewidth}
        $f$ = 0.01, $\alpha = 0.02 $& 76.9 & - & 0.06 & -215 \\
        \hspace{1.17cm}$\alpha = 0.2 $ &61.2 & 0.52 &0.35  & -7  \\
        \hspace{1.17cm}$\alpha = 0.9 $ &49.4 & 0.57 &0.46 & 1  \\
        \Xhline{3\arrayrulewidth}
        \hspace{0.1cm}$f$ = 0.1, $\alpha = 0.02 $&65.9 & 0.52 & 0.28 & 43 \\
        \hspace{1.17cm}$\alpha = 0.2 $ &56.5 & 0.35 &0.69  & 2  \\
        \hspace{1.17cm}$\alpha = 0.9 $ &38.8 & 0.36 &0.78 & -1 \\ 
        \Xhline{3\arrayrulewidth}
        $f$ = 0.25, $\alpha = 0.02 $&63.5 & 0.54 &0.36 & -11 \\
        \hspace{1.17cm}$\alpha = 0.2 $ &52.5 & 0.31 &0.78  & 2 \\
        \hspace{1.17cm}$\alpha = 0.9 $ & 32.6 & 0.31 &0.88 & 1 \\
        \Xhline{3\arrayrulewidth}
        $f = 0$, $\alpha = 0$ (Reference) & 82.8& -  &0.04 & 165\,344
    \end{tabular}
    \caption{Mean waiting time ($\tau$), waiting time break ($W_{\textrm{b}}/\tau$), Pearson coefficient of energy correlation (r), and time lag of the strongest correlation for the nine main simulations of Section \ref{sec:results}.}
    \label{tab:metrics_table}
\end{table}

    \subsection{Total energy, peak energy, and avalanche duration}
    
    \par
\begin{figure}[h!]
        \centering
        \includegraphics[width=\textwidth]{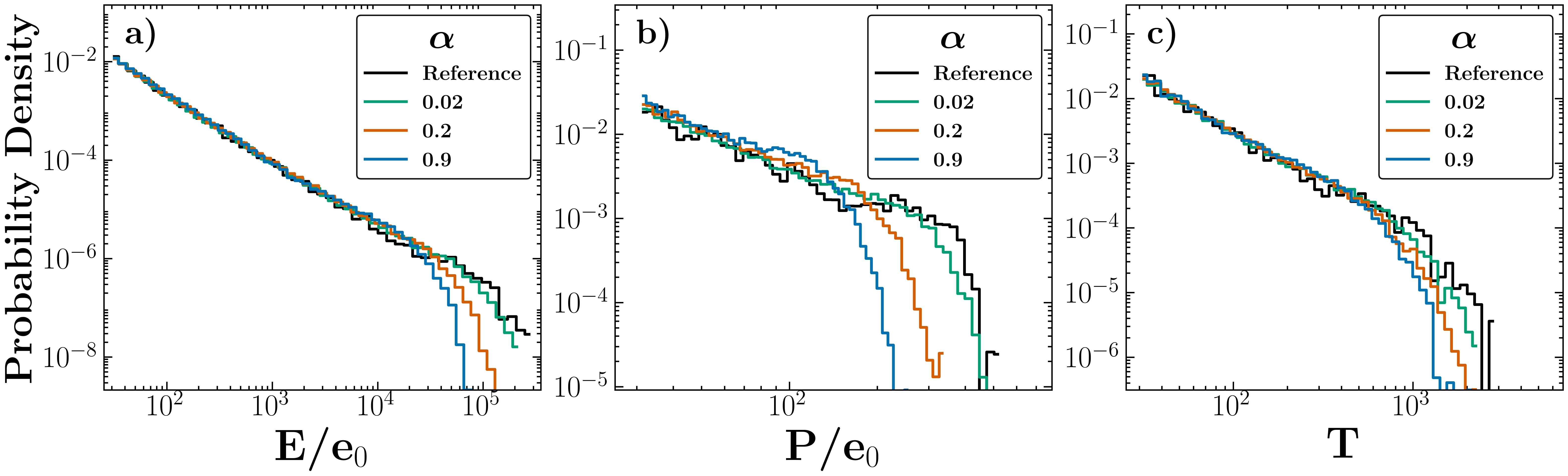}
        \caption{Distributions of total energy (a), peak energy (b), and avalanche duration (c), with $\alpha$ values of 0.02 (green), 0.2 (orange), and 0.9 (blue). We use $f = 0.1$ for all simulations. The black distribution represents a reference simulation, where no transfer occurs between the lattices. Lattices have linear size $N = 48$ and $Z_c = 10$.}
        \label{fig:EPT}
    \end{figure}

 The distributions of typical avalanche size measures, i.e., total energy released by each avalanche ($E$), peak energy released ($P$) and duration ($T$), are shown in Figure \ref{fig:EPT}, panels (a), (b) and (c) respectively. Simulations use $\alpha$ values of 0.02 (green), 0.2 (orange), and 0.9 (blue) and $f = 0.1$. The black distribution represents a reference simulation without transfer between lattices. $E$ and $P$ are normalized by the energy quantum unit $e_{0}$ from Equation (\ref{eqn:quantum}). For small values of $E$, $P$ and $T$, the distributions largely follow the same power-law behavior. However, deviations emerge at larger values, where a stronger connectivity (larger $\alpha$) causes the distributions to drop off sooner. This can be explained by the fact that in strongly connected lattices, large avalanches are quenched by the transfer of nodal variable, resulting in shorter and less energetic avalanches. Note that the state of SOC is preserved even in a regime of strong connectivity ($f = 0.25$), although the distribution of peak avalanche energy starts to lose its power-law form. In comparison, \cite{MacKinnon1997} found that in their model, the SOC state could be disrupted if the non-local effects within a lattice were too significant. Quantitatively, this happened when each avalanching node could influence six remote nodes in the lattice, which corresponds to non-local effects far more important than in our model.
    
    \subsection{Energy Correlation}\label{sec:energy}

   \par
\begin{figure}[h!]
            \centering
            \includegraphics[width=\textwidth]{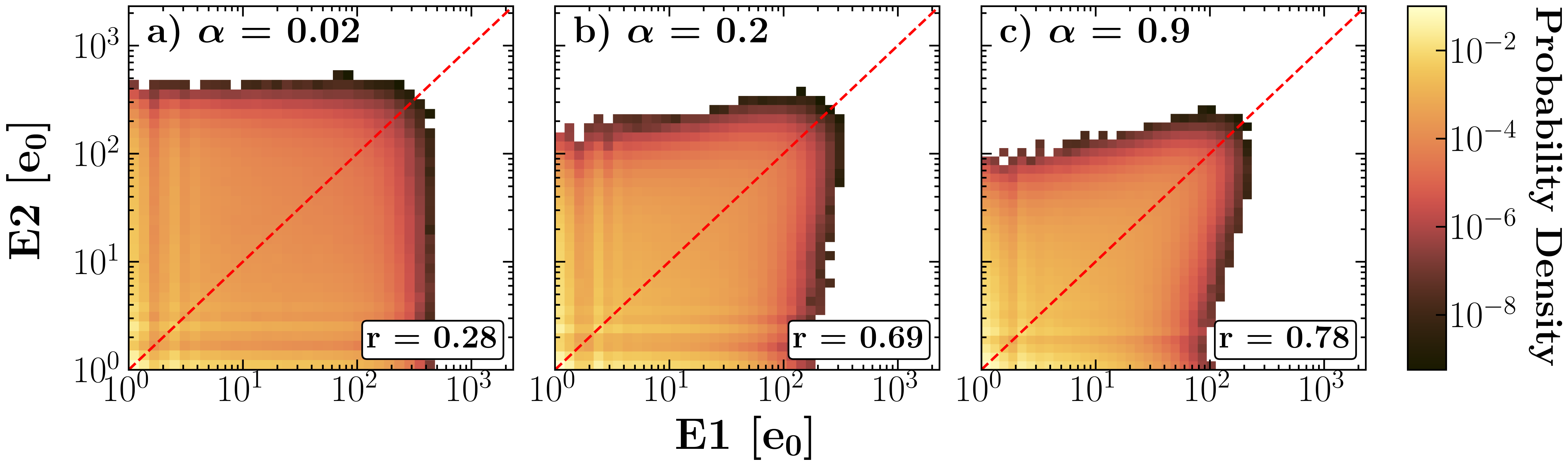}
            \caption{Comparison between the instantaneous energy released of each lattice for $\alpha$ values of 0.02 (a), 0.2 (b), and 0.9 (c). The colormap indicates the probability density. We use $f = 0.1$ for all simulations. Lattices have linear size $N = 48$ and $Z_c = 10$. The dashed red line indicates the 1:1 correlation and the Pearson correlation coefficient $r$ is written in the black box.}
            \label{fig:Ecorr}
        \end{figure}

 Figure \ref{fig:Ecorr} presents two-dimensional histograms of the instantaneous energy released by one lattice versus the other, for $\alpha$ = 0.02 (a), 0.2 (b) and 0.9 (c) with $f = 0.1$. The colormap represents the probability density, while the dashed red line marks the 1:1 correlation. The Pearson correlation coefficient, $r$, is given in the black box. For weak connectivity (panel a), the energies do not correlate, while a strong connectivity (panel c), results in a significant correlation at higher energies, with a Pearson coefficient of $r = 0.78$. This trend is also observed with increasing connectivity, where the strongest correlation occurs for $f = 0.25$ and $\alpha = 0.9$, giving a Pearson coefficient of $r = 0.88$. The Pearson coefficients for all nine main simulations are summarized in Table \ref{tab:metrics_table}. These results suggest that in a regime of strong connectivity, large avalanches in one lattice tend to consistently trigger large avalanches in the other lattice, yet the SOC state is maintained even for parameters as high as $f = 0.75$ and $\alpha = 0.9$ (not shown here, but parameter values used in Section \ref{sec:comparison_sdo}). Note that the horizontal and vertical lines visible at low energies arise from the lattice granularity at small scale, which results in a discrete energy release in units of $e_{0}$, and the finite bin size of the histograms. 
   
    \subsection{Temporal Correlation}\label{sec:time_lag}

     \begin{figure}[h!]
            \centering
            \includegraphics[width=1\textwidth]{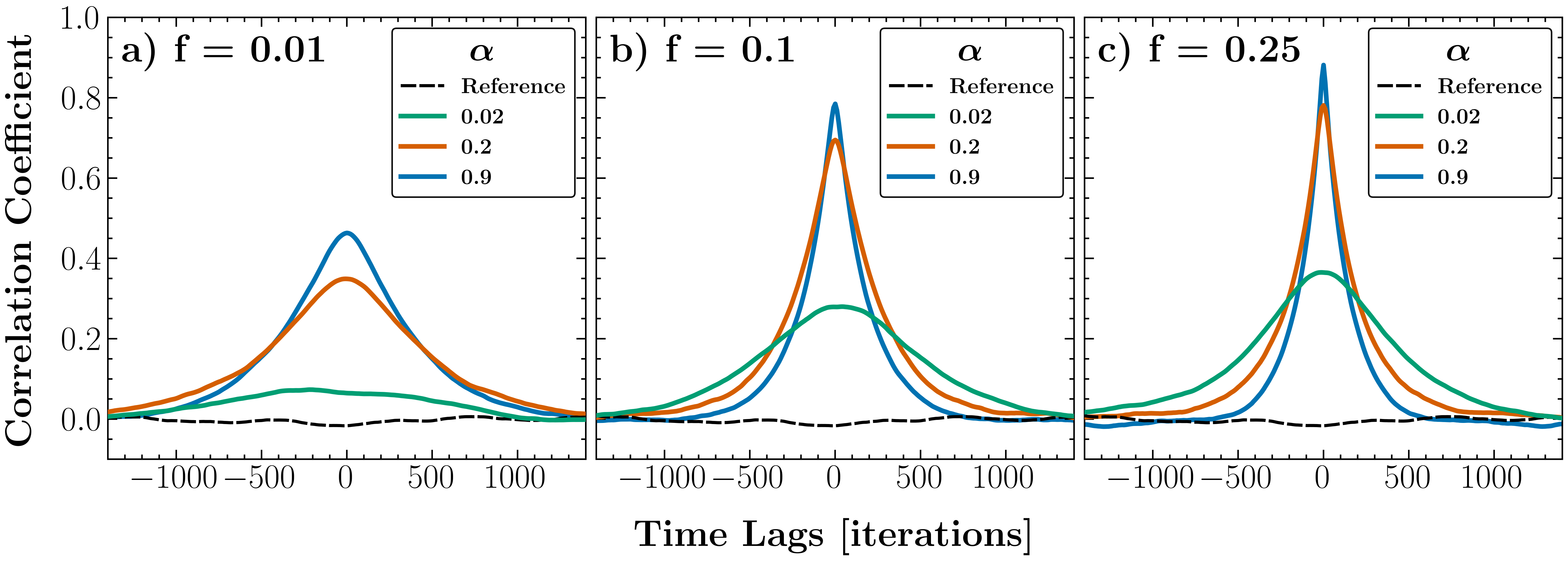}
            \caption{Coefficient of correlation between the two lattices' time series as a function of the time lag, for $f = 0.01$ (a), $f = 0.1$ (b), and $f = 0.25$ (c), with $\alpha$ values of 0.02 (green), 0.2 (orange), and 0.9 (blue). The dashed black line represents a reference simulation, where no transfer occurs between the lattices ($\alpha = 0$).}
            \label{fig:time_lag}
        \end{figure}

    \par The energy release time series of the lattices can be shifted in time with respect to each other to identify the time lag that gives the strongest energy correlation. This is shown in Figure \ref{fig:time_lag} for $f = 0.01$ (a), 0.1 (b), and 0.25 (c), with $\alpha$ values of 0.02 (green), 0.2 (orange), and 0.9 (blue). The dashed black line represents a reference simulation, where no transfer occurs between the lattices. Just like in Figure \ref{fig:Ecorr}, the correlation coefficient increases with stronger connectivity, compared to the reference simulation where the lattices are completely uncorrelated. The shift corresponding to the strongest correlation is summarized in the last column of Table \ref{tab:metrics_table}. As the connectivity strength increases, the temporal correlation between the lattices becomes more important. In the case of $\alpha = 0.9$, the time lag of highest correlation is $\pm \,1$ iteration, meaning that a large energy release in one lattice is will impact the second lattice in the next temporal iteration, due to the substantial transfer of nodal variable.

    \section{Comparison of solar Flares and Avalanches Statistics}\label{sec:comparison_sdo}

        In this section, we introduce a greater range of parameters for our simulations to make comparisons with solar flares observed by the Atmospheric Imaging Assembly \citep[AIA;][]{Lemen2012} instrument on board the Solar Dynamics Observatory \citep[SDO;][]{Pesnel2012}, RHESSI \citep{lin2002}, and the Spectrometer Telescope for Imaging X-rays \citep[STIX;][]{Krucker2020} on board Solar Orbiter \citep[][]{Muller2020}.
        
      \par
\begin{figure}[h!]
            \centering
            \includegraphics[width=0.8\textwidth]{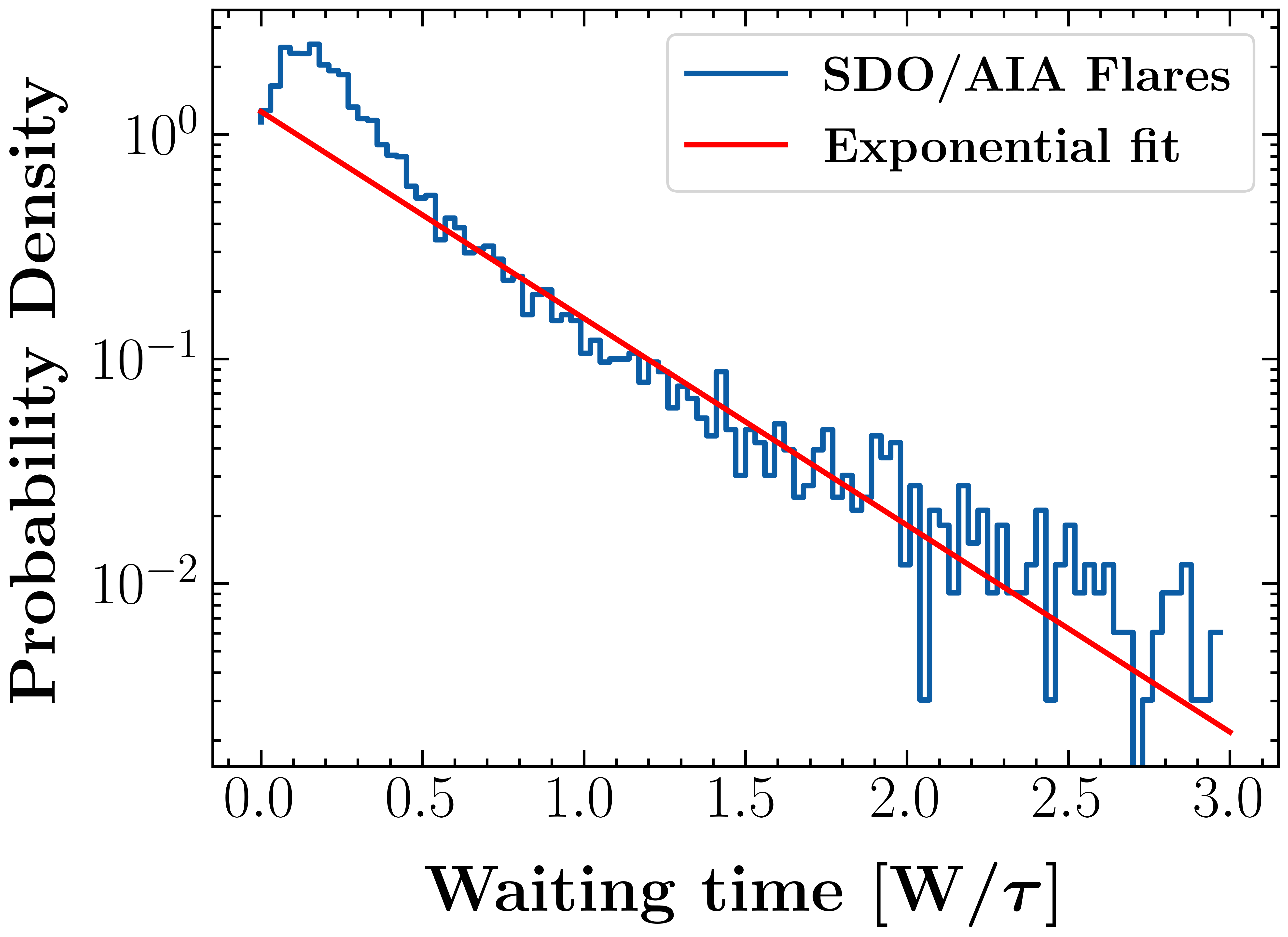}
            \caption{Waiting time distribution of flares observed by the Atmospheric Imaging Assembly (AIA) instrument on board the Solar Dynamics Observatory (SDO). The flare list comes from \cite{Guité2025}. The waiting times are normalized by the arithmetic mean waiting time $\tau$ and the red line indicates an exponential fit.}
            \label{fig:sdo}
        \end{figure}

 Recent investigations by \cite{Guité2025}, using observations from SDO/AIA, RHESSI, and Solar Orbiter/STIX, revealed statistical signatures of sympathetic solar flares, highlighting a tendency for these events to occur at a characteristic angular separation ($\approx 30^{\circ}$ in longitude) and within a waiting time of less than 1.5 hours. In avalanche models, however, angular separation between avalanches on the lattice cannot be defined. By compressing the two-dimensional distribution from Figure 4 in \cite{Guité2025} along the spatial component, we obtain the total waiting time distribution, independent of angular separation, as shown here in Figure \ref{fig:sdo}. The waiting times are normalized by their arithmetic mean waiting time $\tau$, and the red line represents an exponential fit calculated for $W/\tau > 1$. For conciseness, Figure \ref{fig:sdo} presents only the waiting time distribution from SDO/AIA, which includes over 11,000 flares. However, the results from the other two instruments will also be discussed later. For a comprehensive description of the flare datasets and their analysis, we refer readers to \cite{Guité2025}.

      \par
\begin{figure}[h!]
            \centering
            \includegraphics[width=\textwidth]{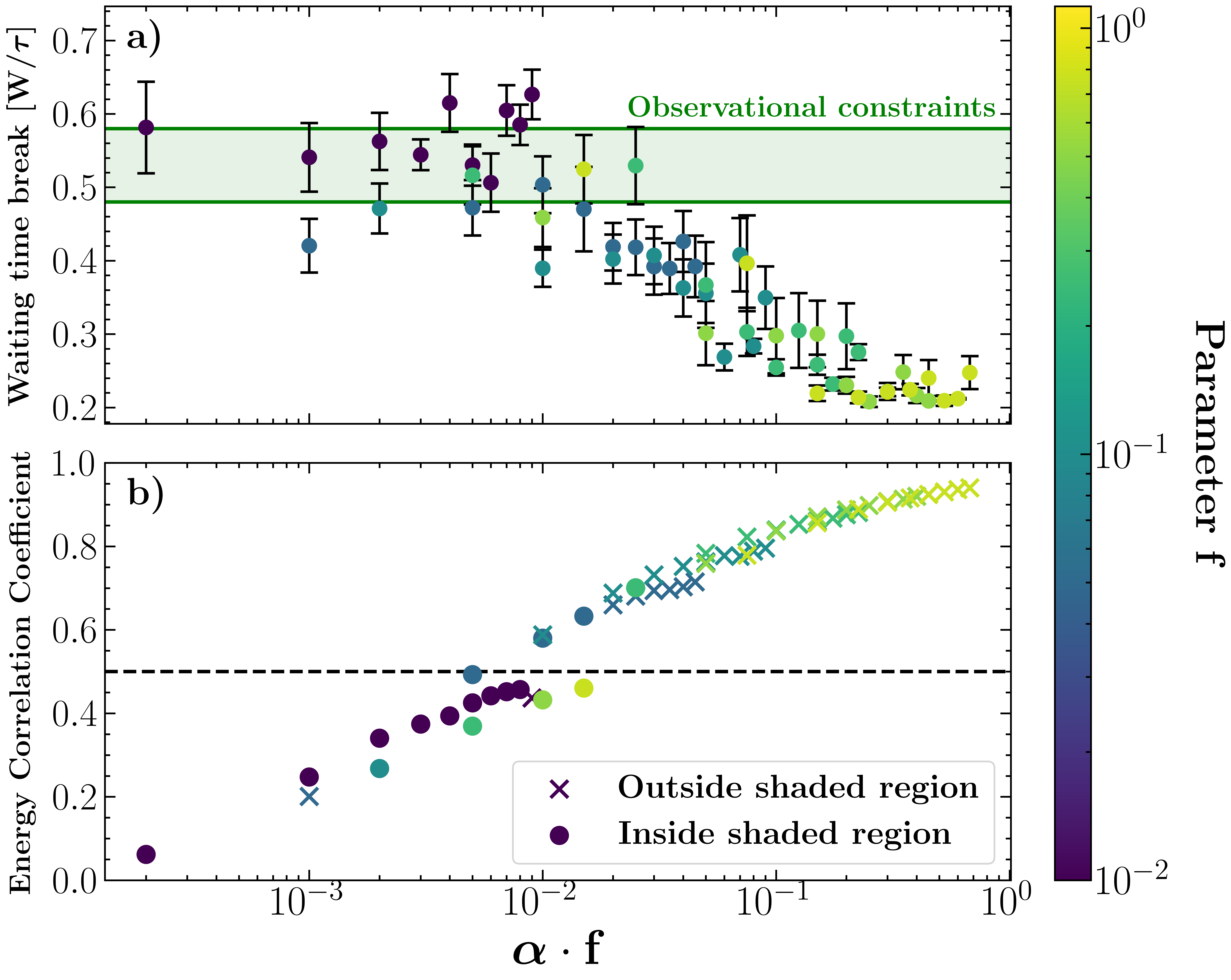}
            \caption{Waiting time break (a) and Pearson coefficient of energy correlation between the two lattices (b) as a function of $\alpha \cdot f$. For each parameter set, 10 simulations are performed to compute the mean values $W_{\textrm{b}}$ and $r$, with the uncertainty represented by the standard deviation across the simulations. Note that no uncertainty is shown in panel (b), since it is too small to be easily visible. The color scale corresponds to the value of $f$. In panel (a) the horizontal lines bounding the shaded green area delimitate the observational range of $W_{\textrm{b}}$, as measured from the waiting time distributions of SDO/AIA, RHESSI, and Solar Orbiter/STIX. In panel (b), points marked as $\bullet$ indicate simulations within the green shaded area of panel (a), while points marked as $\times$ fall outside this region.}
            \label{fig:ensemble}
        \end{figure}

 Similar to Figure \ref{fig:W}, the waiting time distribution of the SDO/AIA flares exhibits an exponential tail, with deviations at shorter waiting times. Interestingly, this deviation also occurs around $W_{\textrm{b}} \approx \frac{1}{2}\tau$ (viz. Figure \ref{fig:exponential_fit}),
      which allows us to constrain the range of $f$ and $\alpha$ values that produce a waiting time break at this threshold in avalanche models. To investigate this, we consider an expanded parameter space, exploring $\alpha = [0.02, 0.1, 0.2, 0.3, 0.4, 0.5, 0.6, 0.7, 0.8, 0.9]$ and $f = [0.01, 0.05, 0.1, 0.25, 0.5, 0.75]$. For each parameter combination, we run 10 simulations to calculate the mean value of $W_{\textrm{b}}$ and its error given by the standard deviation. Figure \ref{fig:ensemble}a illustrates the results, where the mean value of $W_{\textrm{b}}$ is plotted as a function of $\alpha \cdot f$. This combined metric provides a single parameter characterizing the degree of connectivity between the lattices. The color scale indicates the value of $f$ and the horizontal lines, with the shaded green area between them, indicate the observational range of $W_{\textrm{b}}$ measured with the method of Section \ref{sec:waiting_time}. The vertical interval of this region corresponds to the minimum and maximum values among the three $W_{\textrm{b}}$ values calculated from the waiting time distributions of the three instruments: SDO/AIA, RHESSI, and Solar Orbiter/STIX. 
      
      Although there is greater scatter at low connectivity, arising from the difficulty in determining a break point when the distribution closely resembles an exponential, we observe a clear trend: higher connectivity leads to smaller values of $W_{\textrm{b}}$ compared to lower connectivity. Notably, when multiple simulations have the same connectivity ($\alpha \cdot f$), the one with a smaller fraction of connected nodes $f$ tends to have larger waiting time breaks. This finding suggests that if magnetic connectivity between distinct active regions drives sympathetic flares, as proposed by \cite{Guité2025}, then the connectivity must be relatively low to replicate the waiting time distributions observed for SDO/AIA, RHESSI, and Solar Orbiter/STIX flares. Indeed, the range of connectivity values that fall within the observational constraints is approximately $\alpha \cdot f \lesssim 0.025$, with a mean fraction of connected nodes of $f_{\textrm{mean}} \approx 0.14$.
      
        For each simulation shown in Figure \ref{fig:ensemble}a, we calculate the Pearson coefficient $r$ for the energy correlation between the lattices, following the method described in Section \ref{sec:energy}. This coefficient is plotted as a function of $\alpha \cdot f$ in Figure \ref{fig:ensemble}b. Points marked as $\bullet$ indicate simulations within the green shaded area of panel (a), while points marked as $\times$ fall outside this region. The results reveal that simulations meeting the observational constraints of panel (a) exhibit lower energy correlations between the lattices, with Pearson coefficients of $r \lesssim 0.5$, as indicated by the horizontal black dashed line. This finding aligns with the observational analyses of \cite{Guité2025}, who reported no significant correlation between the energies of the first and second flares in pairs of sympathetic flares.

\section{Discussion and Conclusion}\label{sec:discussion}

\par In summary, we introduced a novel avalanche model for sympathetic flares, defined by two parameters, $f$ and $\alpha$, which quantify the degree of connectivity between distinct lattices, representing different active regions on the Sun. Using this model, we observed that the waiting time distribution displays a break where it deviates from the typical exponential function. This break shifts to smaller waiting time as connectivity strengthens. In the regime of strong connectivity, avalanches tend to be shorter and less energetic, while the lattices synchronize temporally and exhibit correlations in their instantaneous energy release. Moreover, we find that the connectivity must be low ($\alpha \cdot f \lesssim 0.025$) to reproduce the observational waiting time distributions of SDO/AIA, RHESSI, and Solar Orbiter/STIX. This result suggests that if magnetic connectivity between active regions is responsible for driving sympathetic flares, it must remain limited in magnitude. This low connectivity is associated with uncorrelated avalanche energies, which is consistent with observations of sympathetic flares on the Sun.

In the context of stellar flares, it is known that cool stars can have much stronger magnetic fields as well as much stronger flares than the Sun \citep{Vasilyev2024}. This could imply that active regions on the surface of such stars are more magnetically connected. In our model, this enhanced connectivity would be associated with highly correlated energies between two subsequent flares, meaning that sympathetic stellar flares could exhibit such correlations. Additionally, the waiting time break would be observed at very low values of $W/\tau$. As a result, the correlation in stellar flare energies and the waiting time distribution are two observational signatures of sympathetic stellar flares that could be monitored on other stars to test the model presented in this study.

%%%%%%%%%%%%%%%%%%%%%%%%%%%%%%%%%%%%%%%%%%%%%%%%%%%%%%%%%%%%%%%%%%%%%%%%%%%
%% Appendix
%
%\appendix   

%%%%%%%%%%%%%%%%%%%%%%%%%%%%%%%%%%%%%%%%%%%%%%%%%%%%%%%%%%%%%%%%%%%%%%%%%%%
%% Acknowledgements
%
\begin{acks}
We thank A.S. Brun and A. Finley for discussions on flare lists and flare statistics.
\end{acks}

%% Available additional data environments:
%% required: authorcontribution, fundinginformation, dataavailability
%% optional: materialsavailability, codeavailability
% \begin{authorcontribution}
%
% \end{authorcontribution}
%
\begin{fundinginformation}

L-SG acknowledges the financial support from the Natural Sciences and Engineering Research Council of Canada (NSERC), the Fonds de recherche du Québec - Nature \& Technologies (FRQNT), a merit scholarship from Hydro-Québec, and an International Student Mobility Grant from the Center for Research in Astrophysics of Quebec (CRAQ). AS and L-SG acknowledge support from the European Research Council (ERC) under the European Union’s Horizon 2020 research and innovation programme (grant agreement No 810218 WHOLESUN). AS acknowledges support from the Centre National d’Etudes Spatiales (CNES) Solar Orbiter, the Institut National des Sciences de l’Univers (INSU) via the Programme National Soleil-Terre (PNST), and the French Agence Nationale de la Recherche (ANR) project STORMGENESIS \#ANR-22-CE31-0013-01. PC acknowledges support from NSERC Discovery grant RGPIN-2024-04050. L-SG and PC are members of the CRAQ.

\end{fundinginformation}
%
% \begin{dataavailability}
%
% \end{dataavailability}
%
% \begin{ethics}
% \begin{conflict}
%
% \end{conflict}
% \end{ethics}

%%% %%%%%%%%%%%%%%%%%%%%%%%%%%%%%%%%%%%%%%%%%%%%%%%%%%%%%%%%%%%
%% Bibliography
%
% Using BibTeX
%
%\bibliographystyle{spr-mp-sola}
%\bibliography{BIB.bib}  

%
% Without BibTeX 
 %\begin{thebibliography}{}
% \bibitem[\protect\citeauthoryear{Author}{Year}]{key}
%   <bibliographical entry>
%
% \bibitem[\protect\citeauthoryear{}{}]{}
%   
%  

\end{document}